\documentclass[useAMS,usenatbib]{mn2e}
\usepackage{epsfig}
\usepackage{amsmath}

\newcommand{\be}{\begin{equation}}  
\newcommand{\beq}{\begin{equation}}  
\newcommand{\ba}{\begin{eqnarray}}  
\newcommand{\ee}{\end{equation}}  
\newcommand{\eeq}{\end{equation}}  
\newcommand{\ea}{\end{eqnarray}}

\newcommand{\apj}{ApJ}  
\newcommand{\apjl}{ApJL}  
\newcommand{\mnras}{MNRAS}  
\newcommand{\aj}{AJ}  
\newcommand{\apjs}{ApJS}  
\newcommand{\nat}{{\it Nature}}

\def\lsim{~\rlap{$<$}{\lower 1.0ex\hbox{$\sim$}}}  
  
\def\gsim{~\rlap{$>$}{\lower 1.0ex\hbox{$\sim$}}}  

\title[Reionization and Galaxy Bias]{The Imprint of Cosmic Reionization on  
Galaxy Clustering}  
  
\author[Wyithe \& Loeb]{J. Stuart B. Wyithe$^1$ \& Abraham Loeb$^2$\\$^1$  
School of Physics, University of Melbourne, Parkville, Victoria,  
Australia\\$^2$ Harvard-Smithsonian Center for Astrophysics, 60 Garden St.,  
Cambridge, MA 02138\\Email: swyithe@physics.unimelb.edu.au,  
aloeb@cfa.harvard.edu}

\begin{document}  
  
  
\maketitle  
  
\label{firstpage}  
\begin{abstract}  
  
We consider the effect of reionization on the clustering properties of
galaxy samples at intermediate redshifts ($z\sim 0.3$--$5.5$). Current
models for the reionization of intergalactic hydrogen predict that
overdense regions will be reionized early, thus delaying the build up of
stellar mass in the progenitors of massive lower-redshift galaxies. As a
result, the stellar populations observed in intermediate redshift galaxies
are somewhat younger and hence brighter in overdense regions of the
Universe.  Galaxy surveys would therefore be sensitive to galaxies with a
somewhat lower dark matter mass in overdense regions.  The corresponding
increase in the observed number density of galaxies can be parameterized as
a galaxy bias due to reionization. We model this process using merger trees
combined with a stellar synthesis code. Our model demonstrates that
reionization has a significant effect on the clustering properties of
galaxy samples that are selected based on their star-formation
properties. The bias correction in Lyman-break galaxies (including those in
proposed baryonic oscillation surveys at $z<1$) is at the level of 10--20\%
for a halo mass of $10^{12}M_\odot$, leading to corrections factors of
1.5--2 in the halo mass inferred from measurements of clustering length.
The reionization of helium could also lead to a sharp increase in the
amplitude of the galaxy correlation function at $z\sim3$. We find that the reionization
bias is approximately independent of scale and halo mass. However since the
traditional galaxy bias is mass dependent, the reionization bias becomes
relatively more important for lower mass systems. The correction to the
bias due to reionization is very small in surveys of luminous red galaxies
at $z<1$.
   
\end{abstract}

   
\section{Introduction}  
  
The clustering of galaxies is often used to study the power-spectrum of the
underlying mass distribution (e.g. Tegmark et al. 2006). Since the data
does not reflect the clustering of mass but rather the clustering of
galaxies, a correction factor termed {\it the galaxy bias} must be applied
to its analysis. This bias factor is a mass (and possibly scale) dependent
property of the galaxy population. Alternatively, assuming that the mass
power-spectrum is known, the masses of the galaxies can be inferred from
their bias by comparing the expected clustering of mass with the observed
galaxy clustering. In either case, if one is trying to determine galaxy
mass using clustering or one is trying to determine the underlying
properties of the mass power-spectra from observations of galaxy
clustering, a theoretical understanding of galaxy bias is required. As a
first estimate, the bias can be calculated from linear theory (Mo \&
White~1996; Sheth, Mo \& Tormen~2001). In addition, through comparison with
N-body simulations, various corrections to the bias have also been
calculated (see Eisenstein et al.~2005 for a summary) to allow more
accurate comparisons with improving data. These contributions to the bias
are physical in the sense that they are due only to the properties of the
dark matter, and so they can be computed (albeit via simulation) from first
principles given an input mass power-spectrum.
  
Recently it has been suggested that an inhomogeneous reionization can lead
to a modification of the observed clustering of galaxies. Babich \&
Loeb~(2006) calculated the modulation of the number density of the
lowest-mass galaxies that result from reionization induced variation in the
thermal history among different regions of the IGM. Although they have
found that the expected effect on the galaxy power-spectrum is much larger
than the difference between competing models of inflation, their analysis
did not extend to high mass galaxies to which future surveys will be
sensitive.  Pritchard, Furlanetto \& Kamionkowski~(2007) considered lower
redshifts and more massive galaxies, but used an ad-hoc ansatz that the
overdensity of galaxies is proportional to the underlying radiation field
and concluded that reionization would leave a redshift dependent imprint on
the galaxy power-spectrum at low redshifts that might interfere with
measurements of the baryonic acoustic peak. These papers did not attempt
to compute the coupling between the mass-to-light ratio of massive galaxies
and the large scale environment.  However galaxy surveys produce clustering
statistics for either flux limited surveys, or for volume limited surveys
in a fixed luminosity range. Computation of the effect of reionization on
the mass-to-light ratio of massive galaxies is therefore critical for
comparison with any real survey.
   
The aim of this paper is to estimate the {\it astrophysical} contribution
to the galaxy bias due to the reionization of the intergalactic medium
(IGM). This contribution is model dependent, requiring knowledge of the
baryonic physics in addition to gravity. The reionization of the IGM is
sensitive to the local large-scale overdensity.  In regions that are
overdense, galaxies are over-abundant for two reasons: first because there
is more material per unit volume to make galaxies, and second because
small-scale fluctuations need to be of lower amplitude to form a galaxy
when embedded in a larger-scale overdensity. The first effect will result
in a larger density of ionizing sources. However this larger density will
be compensated by the increased density of gas to be ionized. In addition,
the recombination rate is increased in overdense regions, but this effect
is counteracted by the bias of galaxies in these regions. The process of
reionization also contains several layers of feedback.  Radiative feedback
heats the IGM and results in the suppression of low-mass galaxy formation
(Efstathiou, 1992; Thoul \& Weinberg~1996; Quinn et al.~1996; Dijkstra et
al.~2004). This delays the completion of reionization by lowering the local
star formation rate, but here again the effect is counteracted in overdense
regions by the biased formation of massive galaxies. The radiation feedback
may therefore be more important in low-density regions where small galaxies
contribute more significantly to the ionizing flux.  Wyithe \& Loeb~(2007)
have modeled the density dependent reionization process using a
semi-analytic model that incorporates the features described above, and so
captures the important physical processes. This model demonstrated that
galaxy bias leads to enhanced reionization in overdense regions, so that
overdense regions are reionized first.

We show that this early reionization leads to an additional bias in the
observed clustering at later epochs in addition to that associated with
enhanced structure formation. We find that the correction to the linear
bias due to reionization could be significantly larger than other
corrections that have been previously considered. Moreover, we show that
the bias correction is larger than the uncertainties in current surveys
over a wide range of redshifts between $1\la z\la 5$.

The outline of the paper is as follows. In \S~\ref{reion} and
\S~\ref{Sbias} we describe the effect of reionization on galaxy formation,
and summarize galaxy bias. We then outline the reasons why we would expect
reionization to yield an additional galaxy bias in \S~\ref{reionbias},
before presenting a model to allow quantitative predictions of the effect
(\S~\ref{model}). We then apply our model to surveys for Ly-break galaxies
(\S~\ref{Lybreak}) and surveys to measure baryonic acoustic oscillations
(\S~\ref{bao}). We then discuss some outstanding issues in
\S~\ref{discussion} before summarizing our conclusions in
\S~\ref{conclusions}.  Throughout the paper we adopt the latest set of
cosmological parameters determined by {\it WMAP} (Spergel et al. 2006) for
a flat $\Lambda$CDM universe.

\section{Reionization and observed galaxy formation}  
  
\label{reion}  
  
The dominant effect of reionization on galaxy formation is believed to  
involve radiative feedback which heats the IGM following the reionization  
of a region, and thus results in the suppression of low-mass galaxy  
formation (Efstathiou, 1992; Thoul \& Weinberg~1996; Quinn et al.~1996;  
Dijkstra et al.~2004).  Standard models of the reionization process assume  
a minimum threshold mass for galaxy halos in which cooling and star  
formation occur ($M_{\rm cool}$) within neutral regions of the IGM. In  
ionized regions the minimum halo mass is limited by the Jeans mass (Barkana  
\& Loeb 2001) in an ionized IGM ($M_{\rm ion}$). We assume $M_{\rm cool}$  
to correspond to a virial temperature of $10^4$K, representing the hydrogen  
cooling threshold, and $M_{\rm ion}$ to correspond to a virial temperature  
of $10^5$K, representing the mass below which infall is suppressed from an  
IGM in which hydrogen has been ionized (Dijkstra et al.~2004).  
  
Observations suggest that hydrogen was reionized by stars prior to  
$z\sim6$ (e.g. White et al.~2003; Fan et al.~2006). However models of  
HeIII reionization suggest that it was the rise of quasars (with  
harder spectra) that resulted in the overlap of HeIII regions at a  
redshift of $z\sim3.5$ (e.g. Wyithe \& Loeb~2003; Sokasian et  
al.~2003). This prediction is consistent with observations that show  
transmission just blueward of the helium Ly$\alpha$ line at $z\sim3$  
(Jacobsen et al 1994; Tytler 1995; Davidsen et al. 1996; Hogan et  
al. 1997; Reimers et al. 1997; Heap et al. 2000; Kriss et al. 2001;  
Smette et al. 2002). In addition, the double reionization of helium  
results in the temperature of the IGM being approximately doubled  
(Schaye et al. 2000; Theuns et al.~2002,2002b).  Thus we assume the  
IGM temperature to change from $T_{\rm IGM}\sim10^4$K to $T_{\rm  
IGM}\sim2\times10^4$K between $z\sim4$ and $z\sim3$.  Calculation of  
the accretion of baryons from an adiabatically expanding IGM into a  
dark matter potential well show that the minimum virial temperature  
for significant accretion is proportional to the temperature of the  
IGM (Barkana \& Loeb~2001). Thus, when helium is reionized at  
$z\sim3.5$, the value of $T_{\rm min}$ is doubled from $T_{\rm ion}$  
to $2T_{\rm ion}$. When considering Helium reionization we assume a  
sudden heating ($\Delta z\la0.1$). However we note that the period of  
heating could be more prolonged.

\section{Galaxy Bias}  
  
\label{Sbias}  
  
Strong clustering of massive galaxies in overdense regions implies  
that these sources trace the higher density regions of IGM.  The  
clustering of galaxies is driven by two effects. The first effect is  
the underlying clustering of the density field.  This clustering may   
be expressed via the mass correlation  
function between regions of mass $M_1$ and $M_2$, separated by a  
comoving distance $R$ is (see Scannapieco \& Barkana~2002 and  
references therein)   
\begin{eqnarray} \nonumber \xi_{\rm  
m}(M_1,M_2,R)&=& \frac{1}{2\pi^2}\int dk k^2 P(k) \\  
&\times&\frac{\sin(kR)}{kR}W(kR_1)W(kR_2), \end{eqnarray}   
where  
\begin{equation} R_{1,2} = \left(\frac{3 M_{\rm 1,2}}{4\pi\rho_{\rm  
m}}\right)^{1/3},   
\end{equation}   
$W$ is the window function (top-hat  
in real space), $P(k)$ the power spectrum and $\rho_{\rm m}$ is the  
cosmic mass density.  The dark-matter halo correlation function for  
halos of mass $M$ is obtained from the product of the mass correlation  
function $\xi_{\rm m}(M,M,R)$ and the square of the ratio between the  
variances of the halo and mass distributions.  This ratio, $b$, is  
defined as the halo bias.  This bias has been discussed extensively  
in the literature, (e.g. Mo \& White~1996; Sheth, Mo \& Tormen~2001).   
However we briefly describe a likelihood based  
interpretation which allows the effects of reionization to be included  
in a natural way.  
  
To see the origin of bias due to enhanced galaxy formation in overdense regions, consider the likelihood (which is proportional  
to the local number density of galaxies) of observing a galaxy at a random location. Given a large scale overdensity $\delta$ of comoving radius $R$,  
the likelihood of observing a galaxy may be estimated from the  
Sheth-Tormen~(2002) mass function as  
\begin{equation}  
\label{LH}  
\mathcal{L}_{\rm g}(\delta) = \frac{(1+\delta)\nu(1+\nu^{-2p}) e^{-a\nu^2/2}}{\bar{\nu}(1+\bar{\nu}^{-2p})e^{-a\bar{\nu}^2/2}},   
\end{equation}  
where $\nu=(\delta_{\rm c}-\delta)/[\sigma(R)]$, $\bar{\nu} = \delta_{\rm  
c}/[\sigma(R)]$ and $\delta_{\rm c}\approx 1.69$ is the critical linear  
overdensity for collapse to a bound object. Here $\sigma(R)$ is the  
variance of the density field smoothed with a top-hat window on a  
scale $R$ at redshift $z$, and $a=0.707$ and $p=0.3$ are constants. Note  
that here as elsewhere in this paper we work with over-densities and  
variances computed at the redshift of interest (i.e. not extrapolated to  
$z=0$). Equation~(\ref{LH}) is simply the ratio of the number density of  
halos in a region of overdensity $\delta$ to the number density of halos  
in the background universe. This ratio has been used to derive the bias for  
small values of $\delta$ (Mo \& White~1996; Sheth, Mo \& Tormen~2001). For  
example, in the Press-Schechter~(1974) formalism we write  
\begin{eqnarray}  
\label{biasPS}  
\nonumber \mathcal{L}_{\rm g}(\delta) &=&  
(1+\delta)\left[\frac{dn}{dM}(\bar{\nu}) +  
\frac{d^2n}{dMd\nu}(\nu)\frac{d\nu}{d\delta}\delta\right]\left[\frac{dn}{dM}(\bar{\nu})\right]^{-1}  
\\&\sim&  
1+\delta\left(1+\frac{\nu^2-1}{\sigma(M)\nu}\right)\equiv1+\delta b_{\rm g},  
\end{eqnarray}  
where $(dn/dM)(\bar{\nu})$ and $(dn/dM)(\nu)$ are the average and perturbed mass functions, and $b_{\rm g}$ is defined as the bias factor.

The observed overdensity of galaxies is
$\delta_{\rm gal}=4/3\times b_{\rm g}(M,z)\delta$, where $b_{\rm g}(M, z)$ is the galaxy  
bias, and the pre-factor of 4/3 arises from a spherical average over the  
infall peculiar velocities (Kaiser 1987).  The value of bias $b_{\rm g}$ for a halo  
mass $M$ may be better approximated using the Press-Schechter formalism (Mo \&  
White~1996), modified to include non-spherical collapse (Sheth, Mo \&  
Tormen~2001)  
\begin{eqnarray}  
\label{bias}  
\nonumber  
b_{\rm g}(M,z) = 1 &+& \frac{1}{\delta_{\rm c}}\left[\nu^{\prime2}+b\nu^{\prime2(1-c)}\right.\\  
&&\hspace{10mm}-\left.\frac{\nu^{\prime2c}/\sqrt{a}}{\nu^{\prime2c}+b(1-c)(1-c/2)}\right],  
\end{eqnarray}  
where $\nu\equiv {\delta_{\rm c}^2}/{\sigma^2(M)}$,  
$\nu^\prime\equiv\sqrt{a}\nu$, $a=0.707$, $b=0.5$ and $c=0.6$. Here $\sigma(M)$ is the variance of the density field smoothed on a mass scale $M$ at redshift $z$. This  
expression yields an accurate approximation to the halo bias determined  
from N-body simulations (Sheth, Mo \& Tormen~2001). Note that in linear theory the bias (equations~\ref{biasPS} and \ref{bias}) is a function of halo mass, but not of overdensity or scale.

\section{Reionization induced galaxy bias in observed galaxy  samples}

\label{reionbias}  
  
We next introduce an additional galaxy bias due to reionization.  In  
addition to color selection criteria, clustering surveys typically  
consider galaxies that are either selected to be above a minimum flux  
threshold (e.g. Adelberger et al.~2005) or to lie in a particular absolute  
magnitude range (e.g. Eisenstein et al.~2005). Suppose that reionization  
caused an overdensity dependent change in the flux per unit halo mass by a  
factor $\mu$. In either of the selection scenarios mentioned above, this  
effect will result in the host halos of survey galaxies being smaller in  
that region by an average factor $\mu$. Following the previous formalism,  
we find the likelihood for observing a galaxy which is subject to a  
decrease in mass-to-light ratio of $\mu$ in a region of overdensity  
$\delta$,  
\begin{eqnarray}  
\nonumber \mathcal{L}_{\rm reion}(\delta) &=&  
\left[(1+\delta)\frac{dn}{dM}(\nu_\mu)\right]\left[(1+\delta)\frac{dn}{dM}(\nu)\right]^{-1}  
\\&\equiv&1+\delta b_{\rm reion}(\delta),  
\end{eqnarray}  
where $(dn/dM)(\nu_\mu)$ is the perturbed mass function evaluated at  
$M/\mu$, and $b_{\rm reion}(\delta)$ is defined to be the bias factor due  
to reionization and which could be a function of $\delta$. Note that since  
surveys are magnitude limited or measured in logarithmic bins of  
luminosity, there is no factor of $\mu^{-1}$ as would be required when  
discussing the number-counts per unit luminosity\footnote{There is also no  
factor of $\mu^{-1}$ to account for depletion as would be appropriate if  
the enhancement in flux were due to gravitational lensing.}.

We may then write an expression for the likelihood of observing a galaxy  
that includes both the bias due to enhanced formation in overdense regions,  
and a possible effect of reionization  
\begin{eqnarray}  
\nonumber  
\label{rbias}  
\mathcal{L}(\delta) &=& \mathcal{L}_{\rm g}(\delta)\times\mathcal{L}_{\rm reion}(\delta)\\  
\nonumber  
    &=&(1+b_{\rm g}\delta)\times(1+b_{\rm reion}\delta) \\  
    &\sim& 1+[b_{\rm g}+b_{\rm reion}(\delta)]\delta.  
\end{eqnarray}   
In the second equality we have parameterized the effect of variance in the  
reionization redshift is an additive contribution to the galaxy bias, and  
have then noted (in the third equality) that we are working in a regime where  
$b\times\delta\ll1$. In the next section we will develop a model that will  
allow us to estimate the magnitude of this effect.

\section{Patchy Reionization and Galaxy Bias}  
  
\label{model}  
  
In this section we describe a model for the effect of reionization on galaxy bias,  
and show that reionization increases the bias in the observed  
overdensity of galaxies relative to the underlying density field. In later sections 
we will use this model to make  
qualitative predictions for the impact of reionization on observed  
clustering in a range of galaxy samples. Our intention is not to  
produce a detailed model in order to make quantitative predictions or  
comparisons with the data. Such a model would require detailed  
numerical simulations, and would in any case require a number of  
uncertain astrophysical assumptions. However our model is adequate for  
the purposes of assessing the importance of reionization in clustering  
measurements, and for making qualitative predictions about its  
dependence on quantities such as survey redshift and luminosity.

\subsection{Reionization redshift and large scale overdensity}  
  
Large-scale inhomogeneity in the cosmic density field  
leads to structure-formation that is enhanced in overdense regions and  
delayed in under-dense regions. Thus, overlap of ionized regions and hence  
heating of the IGM would have occurred at different times in different  
regions due to the cosmic scatter in the process of structure  
formation within finite spatial volumes (Barkana \& Loeb~2004). The  
reionization of hydrogen would have been completed within a region of  
comoving radius $R$ when the fraction of mass incorporated into collapsed  
objects in that region attained a certain critical value, corresponding to  
a threshold number of ionizing photons emitted per baryon  
The ionization state of a region is governed by the  
enclosed ionizing luminosity, by its overdensity, and by dense pockets of  
neutral gas that are self shielding to ionizing radiation.  There is an  
offset (Barkana \& Loeb~2004) $\delta z$ between the redshift when a region  
of mean overdensity $\delta$ achieves this critical collapsed fraction,  
and the redshift ${\bar z}$ when the universe achieves the same collapsed  
fraction on average.  This offset may be computed (Barkana \& Loeb~2004)  
from the expression for the collapsed fraction (Bond et al.~1991) $F_{\rm  
col}$ within a region of overdensity $\delta$ on a comoving scale $R$,  
\begin{equation}  
F_{\rm col}(M_{\rm min})=\mbox{erfc}\left[\frac{\delta_{\rm  
c}-\delta}{\sqrt{2[\sigma_{\rm R_{\rm min}}^2-\sigma_{\rm  
R}^2]}}\right],  
\end{equation}  
yielding  
\begin{equation}  
\label{scatter1}  
\frac{\delta  
z}{(1+\bar{z})}=\frac{\delta}{\delta_{\rm  
c}}-\left[1-\sqrt{1-\frac{\sigma_{\rm R}^2}{\sigma_{\rm R_{\rm  
min}}^2}}\right],  
\end{equation}  
where $\sigma_{\rm R}$ and $\sigma_{R_{\rm min}}$ are the variances in the
power-spectrum at $z$ on comoving scales corresponding to the region of
interest and to the minimum galaxy mass $M_{\rm min}$, respectively. On
large scales equation~(\ref{scatter1}) reduces to
\begin{equation}  
\label{scatter}  
\delta z \approx (1+z)\frac{\delta}{\delta_{\rm c}}  
\end{equation}  
The offset in the ionization redshift of a region depends on its linear  
overdensity, $\delta$. As a result, the distribution of offsets, and  
therefore the scatter in the reionization redshift may be obtained directly  
from the power spectrum of primordial inhomogeneities (Wyithe \&  
Loeb~2004). As can be seen from equation~(\ref{scatter}), larger regions  
have a smaller scatter due to their smaller cosmic variance. Note that  
equation~(\ref{scatter}) is independent of the critical value of the  
collapsed fraction required for reionization. We also note that since at  
high redshift the variance of the linear density field increases  
approximately in proportion to $(1+z)^{-1}$, the typical delay in {\em  
redshift} is almost independent of cosmic time (in addition to not being a  
function of collapsed fraction).  
  
Following the reionization of hydrogen, doubly ionized helium remained in  
the pre-overlap phase. At this time, the mean-free-path of HeIII ionizing  
photons was therefore limited to be smaller than the size of the HeIII  
regions. As is the case for hydrogen, the ionization state of these regions  
was therefore dependent on the local source population. If it is true that  
quasars were responsible for the reionization of helium, then these are  
much rarer sources than the galaxies responsible for the reionization of  
hydrogen. As a result there would be large fluctuations in the HeIII  
reionization redshift due to Poisson fluctuations in the number of sources  
and variations in the opacity of the IGM (Reimers et al. 2006).  These  
fluctuations would not be simply related to the local large scale  
overdensity. On the other hand, the arguments regarding the fluctuations in  
the redshift of hydrogen reionization due to enhanced structure formation  
in overdense regions must also apply to the reionization of helium, and  
these will be present in addition to the Poisson noise. As already  
mentioned, the delay in reionization due to an overdensity $\delta$ is not  
a function of cosmic time. Thus we see from equation~(\ref{scatter}) that  
since the delay is also independent of collapsed fraction (which we expect  
to be different for hydrogen and helium reionization), the delay ($\delta  
z$) in the redshift of HeIII reionization for a particular value of the  
comoving overdensity $\delta$ is equal to the delay for the reionization  
of hydrogen. As a result, in an overdense region both hydrogen and helium  
would be reionized early by the same offset in redshift.  The large-scale  
variations in the reionization redshifts of hydrogen and helium lead to a  
different accretion histories for galaxies, which in-turn lead to different  
star-formation histories, and thus a change in the luminosity of a galaxy  
given a total stellar mass due to the different age distribution of the  
stellar population. 
  
Before proceeding, we draw attention to the approximation of sudden
reionization in which the reionization of a volume on a scale $R$ occurs at
a redshift $z$. Of course some regions within that volume will have been
reionized earlier. However our point is that, on average, a region of IGM will be
reionized earlier by $\Delta z$ within a volume of scale $R$.  A critical
component of our model is the assumption that the average variation in the
redshift at which the gas that ultimately makes the progenitor galaxies was
reionized is also equal to $\Delta z$.
  
Weinmann et al.~(2007) have recently employed numerical simulations of
reionization to compute whether a galaxy observed at the present time
formed in a region of IGM prior to it being reionized, or whether it formed
in a region that had already been reionized. In their work the time of
formation of a galaxy refers to the identification of the earliest
progenitor of the local galaxy above the resolution limit of the
simulation. These authors find that more massive galaxies had progenitors
that formed in neutral regions while less massive galaxies formed in
ionized regions. They also conclude that there is no correlation between
the reionization history of field galaxies and their environment or large
scale clustering (however see discussion below). While very useful in
understanding the relation between the early formation histories of
galaxies and the reionization process, these findings are not directly
applicable to our discussion, which aims to calculate the average effect of
reionization on all the progenitors of a low redshift galaxy rather than on
its earliest progenitor.  

But interestingly, the numerical results presented by Weinmann et
al.~(2007) show consistency with the quantitative expectations of our
simple model. Their Figure~6 shows the relation between the reionization
redshift for a massive galaxy and the local overdensity of massive galaxies
within 10 comoving Mpc. The simulations predict a large scatter of $\pm1$
redshift unit about a mean relation that varies by $\Delta z\sim0.8$
between present-day galaxy overdensities of $-1$ and 1 on 10 comoving Mpc
(cMpc) scales, with overdense environments reionizing at higher
redshift. While the scatter is large as expected, the statistical accuracy
of the measured mean is significantly below $\Delta z=0.8$ due to the large
sample size of model galaxies. This implies that Weinmann et al.~(2007) do
indeed infer a relation between large scale environment and the mean
reionization redshift, but that the variation in the mean relation is not
significant with respect to the scatter among individual galaxies.  We can
compare the mean relation of Weinmann et al.~(2007) for the reionization
redshift of the earliest progenitor from simulations with expectations from
our simple model for patchy reionization. On a scale of 10 cMpc,
the variance in the density field at $z=0$ is
$\sigma(10\mbox{cMpc})\sim0.8$. Since the bias is around $b\sim1.1$ for the
$5\times10^{12}M_\odot$ galaxies from which the simulated relation was
calculated, we expect 1-sigma fluctuations in the overdensity of galaxies
at $z=0$ on 10 cMpc scales to be $\pm1$. Thus the numerical
simulations of Weinmann et al.~(2007) predict fluctuations in the
reionization redshift around a mean of $z\sim8.5$ (predicted by their
model) of $\delta z\sim0.4$ for the earliest progenitor of massive local
galaxies.  Our simple model predicts the fluctuation in the reionization
redshift of the IGM with an overdensity of $\delta\sim0.8$ to be $\delta
z\sim (1+z)\times(\delta D(z))/\delta_{\rm c}\sim0.6$, where $D$ is the
growth factor (from equation~\ref{scatter}). This number is similar to the
typical fluctuations in the reionization redshift of the earliest
progenitor in the simulations of Weinmann et al.~(2007).  Weinmann et
al.~(2007) also argue that the way in which a galaxy is reionized (either
externally or internally) is not sensitive to the local overdensity of
galaxies. Thus numerical simulations predict that the process of
reionization is similar in overdense and underdense regions, but that
reionization is accelerated in overdense regions. These findings from
numerical simulation support the basic assumptions of our simple model.

\subsection{Model of Reionization induced bias}

To develop our model we consider a galaxy residing in a halo of mass  
$M$ at $z\ll z_{\rm reion}$. This galaxy has accreted its mass via a  
merger tree, which we generate using the method described in  
Volonteri, Haardt \& Madau~(2003). We describe this tree as having a  
number $N_{\rm halo}(z_j)$ of halos of mass $M_i(z_j)$ at redshift  
$z_j$, where the number of redshift steps is $N_z$, with values of  
redshift that increase from the redshift of the primary halo in the  
tree so that $z_0=z$. These halos grow in mass due to mergers of progenitor
halos, and due to accretion (which, in the Press-Schechter formalism,  
is the sum of mergers with halos below the resolution limit of the  
merger tree).  
  
First consider halos above the minimum mass for star formation (which is either $M_{\rm cool}$ in neutral regions, or $M_{\rm reion}$ in reionized regions respectively).  
At each redshift step, a fraction of the baryonic mass  gained by these halos through accretion is turned into stars, thus  
\begin{eqnarray}  
\nonumber  
\Delta M_{\star,i}(z_j) &=& f_\star\frac{\Omega_{\rm b}}{\Omega_{\rm m}}\left(M_i(z_j)-M_i(z_{j+1})\right)\hspace{2mm}\mbox{for}\hspace{2mm}M>M_{\rm min}\\  
\Delta M_{\star,i}(z_j) &=& 0 \hspace{5mm}\mbox{otherwise},  
\end{eqnarray}  
where $f_\star$ is the star formation efficiency. We choose $f_\star=0.3$ throughout this paper, though our conclusions are not sensitive to this choice.

In addition, we assume that whenever a progenitor halo $i$ at redshift $z_j$ in the merger tree crosses the minimum mass for star formation through the merger of two sub-units $M_{i,1}(z_{j+1})$ and $M_{i,2}(z_{j+1})$, stellar mass is added in the amount  
\begin{equation}  
\Delta M_{\star,i}(z_j) = \left(f_\star\frac{\Omega_{\rm b}}{\Omega_{\rm m}}M_i(z_j)\right) - \left(M_{\star,i,1}(z_{j+1}) +M_{\star,i,2}(z_{j+1})\right),  
\end{equation}  
where $M_{\star,i,1}(z_{j+1})$ and $M_{\star,i,2}(z_{j+1})$ are the stellar mass content of the progenitors prior to the merger.   
Similarly, whenever a progenitor halo $i$ at redshift $z_j$ in the merger tree crosses the minimum mass for star formation through accretion, stellar mass is added in the amount  
\begin{equation}  
\Delta M_{\star,i}(z_j) = f_\star\frac{\Omega_{\rm b}}{\Omega_{\rm m}}M_i(z_j) - M_{\star,i}(z_{j+1}),  
\end{equation}  
where $M_{\star,i}(z_{j+1})$ is the stellar mass content of the halo at the previous redshift step.   
The subtraction of the second term is necessary in each of the latter cases because the minimum mass in a region increases suddenly at the local reionization epoch.   
The total stellar mass added at each step is the sum of these three contributions. We may then construct a stellar-mass accretion history  
\begin{equation}  
\frac{d(\Delta M_\star)}{dz}(z_j) \sim \frac{1}{z_j-z_{j-1}}\Sigma_{i=0}^{N_{\rm halo}(z_j)}\Delta M_{\star,i}(z_j).  
\end{equation}  
  
Our scheme neglects any star formation that may occur in recycled gas  
following a major merger. However star-formation in recycled gas at  
low redshift is by definition already above the minimum threshold for  
star-formation and so should not be sensitive to the local redshift of  
reionization. We have used a sudden transition of the minimum virial  
temperature for star formation, rather than the more gradual  
transition described by the filtering mass, which takes account of the  
formation time-scale for a collapsing halo and the full thermal  
history of the IGM. In the hierarchical build-up of a halo, the merger  
of collapsed progenitors, combined with accretion of mass onto these  
progenitors can be identified with the formation of the final halo  
within the spherical collapse model and Press-Schechter formalism. The  
sudden transition of virial temperature is therefore the appropriate  
choice for our model since the star formation is calculated to  
coincide with the formation of a halo during a merger tree. The  
fraction of mass in the halo at redshift $z$ that was already  
collapsed prior to reionization is therefore explicitly accounted for  
in our model.

Before proceeding we note that the calculation of the merger tree is independent of the large-scale overdensity $\delta$.   
We demonstrate this explicitly in Appendix~\ref{app1}, for a  
cosmology that ignores the cosmological constant (as is appropriate at  
high redshift). Thus, to estimate the additional contribution to the bias  
that is due to changes in the star formation history  
associated with the reionization variable redshift, we may compute one merger tree  
within the mean background cosmology, and then change only the  
reionization redshift to account for the overdensity of the region in  
which the parent halo formed. The total bias for the observation of  
this galaxy is then the reionization induced bias computed from this  
merger tree, plus the usual galaxy bias.  This independence of the  
merger tree on $\delta$ greatly simplifies calculation of the  
dependence of the apparent brightness of a galaxy within a halo of  
mass $M$ on the large-scale overdensity.

We next compute the spectrum of stellar light that results from this star  
formation history using the stellar population model of Leitherer et  
al.~(1999). We assume a 1/20th solar metallicity population with a  
Scalo~(1998) mass-function and begin with the time dependent spectrum for a  
burst of star formation\footnote{Model spectra of star forming galaxies  
obtained from http://www.stsci.edu/science/starburst99/.}. This yields the  
emitted energy per unit time per unit frequency per solar mass of stars  
$d^2\epsilon_\nu/dtdM(t_0-t_j)$ at a time $t_0-t_j$ following the burst,  
where $t_0$ and $t_j$ are the ages of the universe at the redshift of the  
primary halo ($z_0$) and $z_j$ respectively. The flux (erg/s/cm$^2$/Hz)  
from the galaxy at $z$ then follows from the sum over the starburst  
associated with each star formation episode. We find  
\begin{equation}  
\label{fnu}  
f_\nu = \Sigma_{j=0}^{Nz} \Delta M_\star(z_j) \frac{d^2\epsilon_\nu}{dtdM}(t_0-t_j) \frac{1}{4\pi D_{\rm L}(z_0)^2}(1+z_0),  
\end{equation}  
where $D_{\rm L}$ is the luminosity distance at $z$. We note that our scheme neglects enrichment of gas prior to star formation, so that all star-bursts are assumed to have the same metallicity. Since we are not computing the contribution from star-formation in recycled gas, we do not expect this assumption to have a large influence on our results. Moreover, the UV spectra of galaxies are very sensitive to the dust content of a galaxy and therefore also sensitive to the metallicity. However the reionization bias would be sensitive to the ratio of fluxes in which the effect of the dust on a single spectrum will vanish.  
  
We can then use equation~(\ref{fnu}) to compute the spectra for two star-formation histories that correspond to the mean universe and to an overdensity $\delta$, with reionization redshifts separated by $\delta z$. These spectra in turn allow us to determine ratio of fluxes and hence to compute a value of $\mu(\delta)$. As described above, in order to compute the contribution of reionization to galaxy bias we make this comparison  
directly, using the same merger tree with different values of the  
reionization redshift to obtain different star formation histories.

Equation~(\ref{fnu}) implies that the apparent flux of a galaxy will  
be sensitive to its star-formation history, which will in turn be  
sensitive to the redshift of reionization. We would now like to  
calculate the typical flux change induced by the effect of a large  
scale overdensity on the reionization redshift. In order to achieve  
this we must determine the change in flux as a function of scale, and  
at a range of over-densities. As shown above, the delay in the  
reionization redshift is proportional to the variance of the  
power-spectrum on the scale of interest. Since the variance decreases  
towards large scales, we find that the fluctuations in reionization  
redshift should also be smaller on large scales. In order to  
investigate the scale dependence of the bias, we could therefore  
compute the difference in star formation histories corresponding to  
different delays in reionization over a number of spatial  
scales. However we find that for individual merger trees, the value of  
the apparent magnitude change (i.e. the logarithm of the  
observed flux) is approximately proportional to the delay in  
reionization. Thus we can estimate the change in magnitude due to reionization using a first order expansion in $\delta z$  
\begin{eqnarray}  
\label{deltaf}  
\nonumber  
2.5\log_{10}{\mu} &=& 2.5\frac{d\log_{10}[f_\nu(z_{\rm ol})]}{dz_{\rm ol}}\delta z\\  
               &\sim& 2.5\log_{10}\left[\frac{f_\nu(z_{\rm ol})}{f_\nu(z_{\rm ol}+\Delta z)}\right]\frac{\delta z}{\Delta z},  
\end{eqnarray}  
where $z_{\rm ol}$ is the overlap (reionization) redshift, which we assume  
to be $z=6$ throughout this paper (e.g. White et al.~2003; Fan et  
al.~2006), and $\Delta z=0.25$ is the separation in overlap redshifts of  
the two star formation histories computed for each merger tree.  Thus the  
magnitude change can be computed for a single length scale $R$ and  
overdensity $\delta$, and then translated to other length scales and over  
densities in proportion to the variance. We employ this approximation which  
greatly simplifies our calculations.

Given a scale $R$ and variance $\sigma(R)$ we can now estimate the contribution of reionization to galaxy bias. For each merger tree $k$, we compute the bias averaged over likeli-hoods at each $\delta$ in the density field  
\begin{equation}  
b_{{\rm reion},k} = \frac{1}{\sqrt{2\pi}\sigma(R)}\int d\delta \left[\frac{1-\mathcal{L}(\delta)}{\delta}\right]\exp{\left(-\frac{\delta}{2\sigma(R)}\right)^2}.  
\end{equation}  
To get the average bias for the galaxy population, we then average the bias evaluated using $N_{\rm trees}$ different merger trees,  
\begin{equation}  
b_{\rm reion} = \frac{1}{N_{\rm trees}}\Sigma_{k=0}^{N_{\rm trees}}b_{{\rm reion},k}.  
\end{equation}  
In the remainder of this paper, we use the above model to estimate the contribution of reionization to galaxy bias for several existing and planned galaxy surveys.

\section{Ly-break Galaxies}  
  
\label{Lybreak}  
  
\begin{figure*}  
\includegraphics[width=13cm]{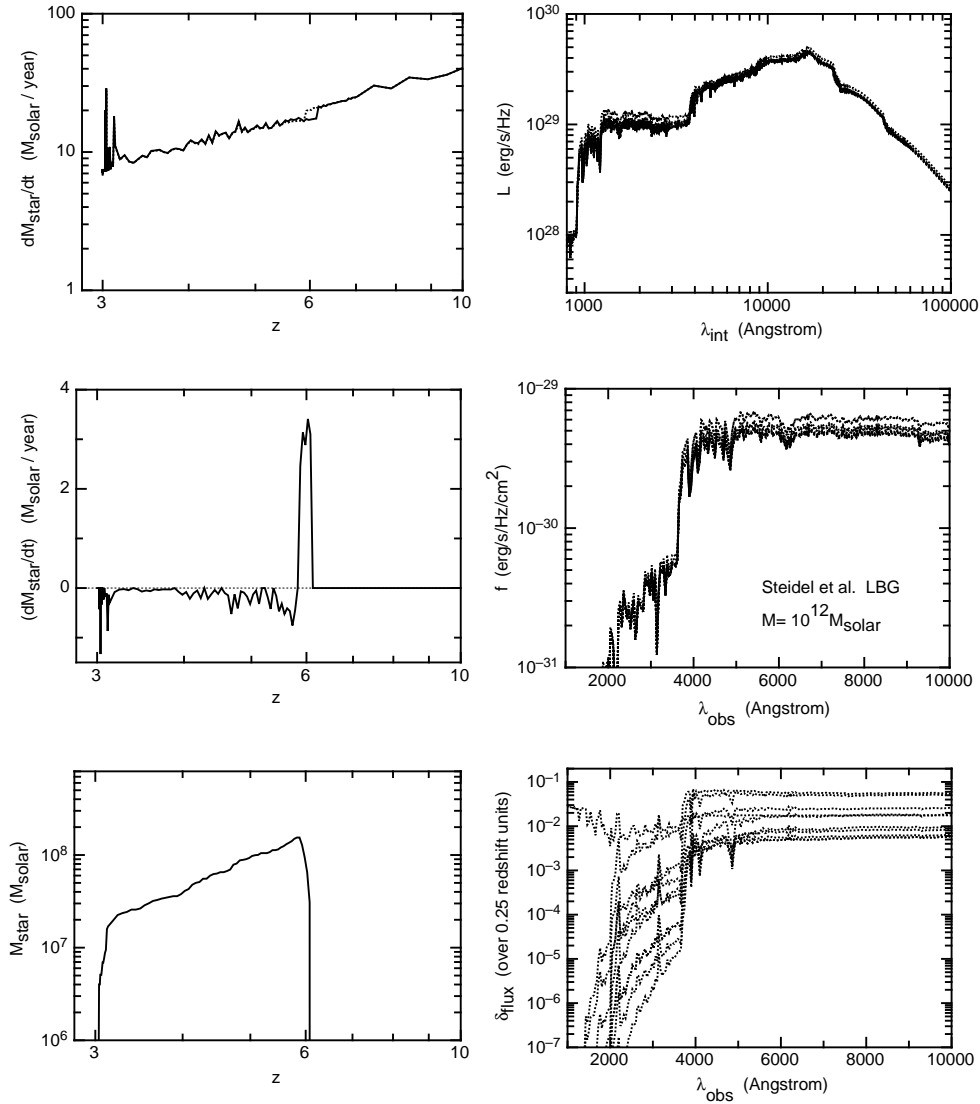}   
\caption{ The effect of reionization on the star formation histories of galaxies. These examples correspond to typical Ly-break galaxies at $z=3$ in the survey of Steidel et al. (2003). \textit{Upper Left:} The star formation rate summing over all halos that end up as part of the galaxy in a $10^{12}M_\odot$ halo at $z=3$. The solid and dotted lines refer to histories where overlap occurred at $z=6.125$ and $z=5.875$ respectively. \textit{Central Left:} The difference in star formation rate for the two histories. \textit{Lower Left:} The difference in cumulative stellar mass for the two histories. \textit{Upper Right:}  Rest frame luminosity of ten example Ly-break galaxies at $z=3$. \textit{Central Right:}  Observed flux (including Ly$\alpha$ absorption) for the ten example galaxies.  \textit{Lower Right:} The magnitude change induced by a delay in reionization of 0.25 units of redshift.  }  
\label{fig1}  
\end{figure*}   
  
\begin{figure*}  
\includegraphics[width=13cm]{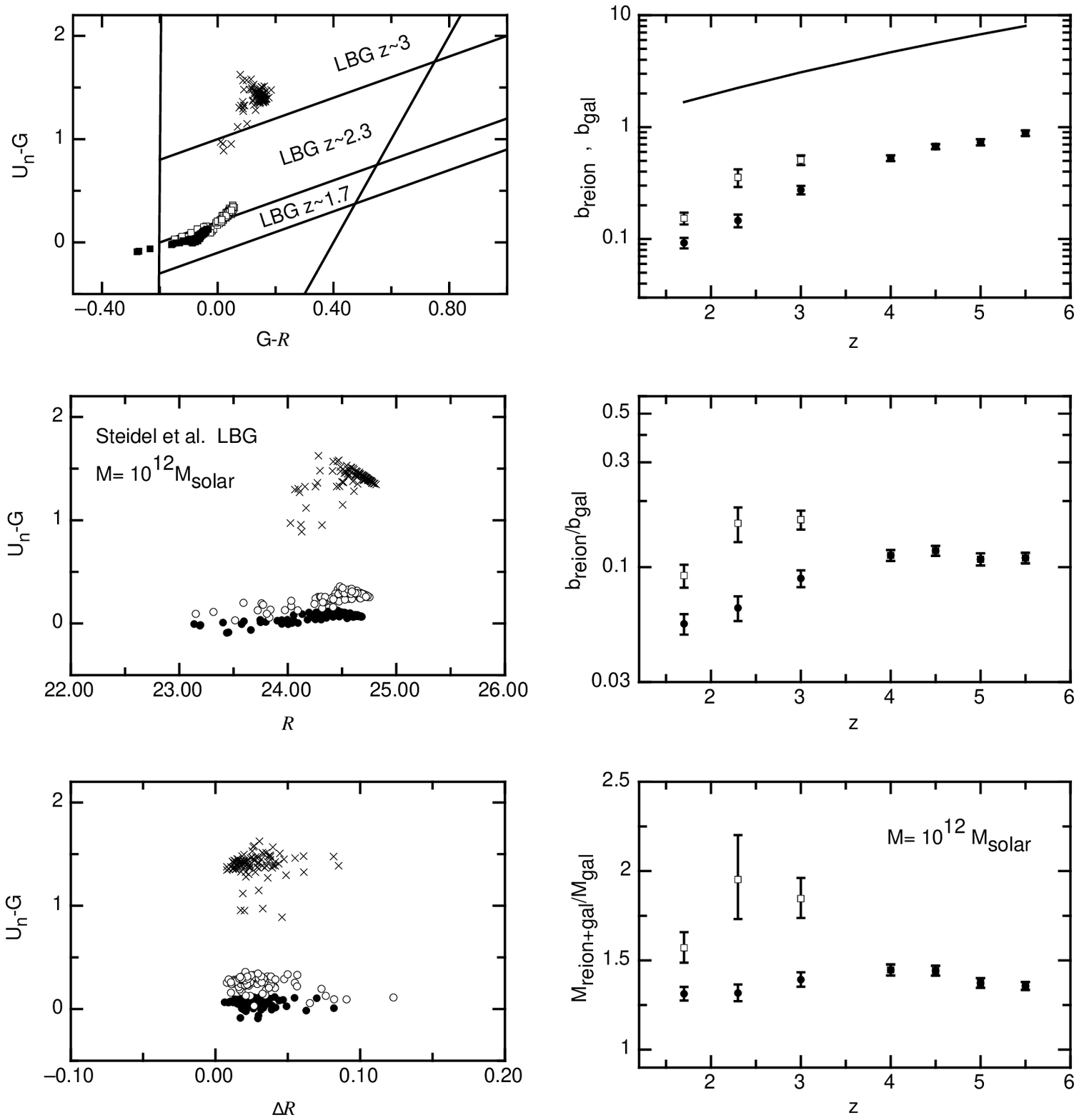}   
\caption{ Examples of clustering bias in Ly-break galaxies induced by  
reionization. \textit{Upper Left:} The primary color selection (Steidel et  
al.~2003) for LBGs at $z\sim1.7$ (solid points), $z\sim2.3$ (open points)  
and $z\sim3$ (crosses). \textit{Central Left:} The apparent magnitudes and  
colors of the model galaxies. \textit{Lower Left:} The $U_n-G$ color as a  
function of the change in $\mathcal{R}$-band magnitude induced by  
reionization. \textit{Upper Right:} The bias introduced by reionization in  
cases where helium reionization at $z\sim3.5$ is considered in addition to  
hydrogen reionization (open squares) and where it is not (solid squares). The bias  
was computed assuming a flux evaluated at a rest-frame wavelength of $1350$\AA within a $400$\AA  
window. The galaxy bias is shown by the solid line for comparison. The  
error bars represent the statistical noise in the simulations due to the  
finite number of merger trees. \textit{Central Right:} The ratio of the  
component of bias introduced through reionization to the usual galaxy  
bias. \textit{Lower Right:} The factor by which the mass will be  
overestimated in clustering analyses where reionization is not  
considered. }  
\label{fig2}  
\end{figure*}

As a first application of our model we construct mock spectra  
corresponding to Ly-break galaxies (Steidel et al. 2003) at  
$z=3$. Examples of the model star formation history and resulting  
galaxy spectra are shown in Figure~\ref{fig1} assuming a halo mass of  
$M=10^{12}M_\odot$ (Adelberger et al.~2005). Here we include only  
hydrogen reionization as the mechanism that introduces fluctuations in  
the star-formation history.  
  
In the upper left panel of Figure~\ref{fig1} we show an example of the star  
formation rate summed over all halos that end up as part of the galaxy in a  
$10^{12}M_\odot$ halo at $z=3$. Here the solid and dotted lines refer to  
histories where the overlap of ionized regions occurred at $z=6.125$ and  
$z=5.875$ respectively. The effect of the reionization redshift on these  
star formation histories is more easily seen in the central left panel  
where we plot the difference in star formation rate between the two  
histories. Figure~\ref{fig1} shows that early reionization initially  
results in a deficit of star formation. This deficit is then made up  
continually until $z=3$. By $z=3$ the total stellar mass is the same for  
both histories ($M_\star=f_\star\Omega_{\rm b}/\Omega_{\rm m}M$) as  
required for consistency of the model. This behavior is also demonstrated in the lower left panel of  
Figure~\ref{fig1} where we plot the difference in cumulative stellar mass,  
summing over all halos that end up as part of the primary galaxy at $z=3$.  
  
Figure~\ref{fig1} also shows examples of the resulting galaxy  
spectrum. In the upper right panel we show the rest frame luminosity  
for ten example $10^{12}M_\odot$ galaxies at $z=3$. In each case  
reionization was at $z=6.125$. The differences between these spectra  
arise due to the slightly different age distribution of the stellar  
populations that result from the stochastic buildup of mass in the  
merger tree.  In the central right panel of Figure~\ref{fig1} we show  
the corresponding observed flux [including mean Ly$\alpha$ absorption, see e.g. Fan et al.~(2006)] for  
the same ten example Ly-break galaxies at $z=3$.  The distinctive  
Ly-break near 4000\AA~ is clearly visible in these spectra. Finally,  
in the lower right hand panel we show the fractional change  
($\delta_{\rm flux}$) in the observed flux that is induced by a delay  
in reionization of $\Delta z=0.25$ units of redshift. This flux  
change, which is related to the parameter $\mu$ through  
equation~(\ref{deltaf}) corresponds to differences in the  
star-formation histories that are comparable to the example shown in  
the left hand panels of Figure~\ref{fig1}. The fluctuations are at the  
level of a few tenths to a few percent. We investigate the bias that  
arises from the resulting values of $\mu$ in \S~\ref{ss_reionbias}.

\subsection{The colors of simulated Ly-break galaxies}  
  
Our aim in this paper is to evaluate the importance of reionization  
with respect to galaxy bias in measurements of galaxy clustering. To  
this end we have constructed model star-formation histories that  
include the effect of reionization, and computed the effect of  
reionization on the corresponding model galaxy spectra. In order for  
our results to be applicable to surveys of real galaxies, we must, at  
a minimum demonstrate that our model produces realistic spectra with  
colors that would see the model galaxies selected into the survey of  
interest. Therefore, before describing our results for the  
reionization induced bias we demonstrate that our model galaxies have  
colors and magnitudes that correspond to those of real Ly-break galaxies (LBG).  
  
In the upper left panel of Figure~\ref{fig2} we show the position of  
100 model Ly-break galaxies within the primary color selection\footnote{To estimate the colors of Ly-break galaxies we assume top-hat filters of central wavelength $\lambda_0$ and width $\Delta \lambda$ to approximate the filter set used in Steidel et al.~(2003). We use $AB$-magnitudes throughout this paper. The filters have $(\lambda_0,\Delta\lambda)= (3550,600)$ for the $U_n$-band; $(\lambda_0,\Delta\lambda)=(4780,1100)$ for the $G$-band; $(\lambda_0,\Delta\lambda)=(6830,1250)$ for the $\mathcal{R}$-band; $(\lambda_0,\Delta\lambda)=(8100,1650)$ for the $I$-band.}   
(Steidel et al.~2003) for LBGs at $z\sim1.7$ (solid points),  
$z\sim2.3$ (open points) and $z\sim3$ (crosses). The galaxies at  
$z\sim3$ are well separated from those at lower redshift due to the  
Ly-break moving to a wavelength beyond the $U_n$-band. Our model  
galaxies at $z\sim1.7$ and 2.3 have similar colors that are close to  
the selection cutoff. This is consistent with the observed galaxies,  
which have overlapping redshift distributions when selected via this  
criteria (Adelberger et al.~2005). For their clustering analysis  
Adelberger et al.~(2005) restricted themselves to objects with  
$23.5<\mathcal{R}<25.5$. In the central left panel we show the position of our  
model galaxies in a color-magnitude diagram. The apparent magnitudes  
of these model galaxies are consistent with the observed  
population. Thus our model produces Ly-break galaxies with both the  
correct colors, and the correct luminosity. Finally we check that  
reionization induced changes in the observed flux are not sensitive to  
the observed galaxy color. In the lower left panel of  
Figure~\ref{fig2} we show the $U_n-G$ color as a function of the change  
in $\mathcal{R}$ magnitude induced by reionization. We find no systematic trend  
of the flux variation with galaxy color.  
  
\subsection{Reionization induced bias for Ly-break galaxies}  
\label{ss_reionbias}  
  
\begin{figure*}  
\includegraphics[width=13cm]{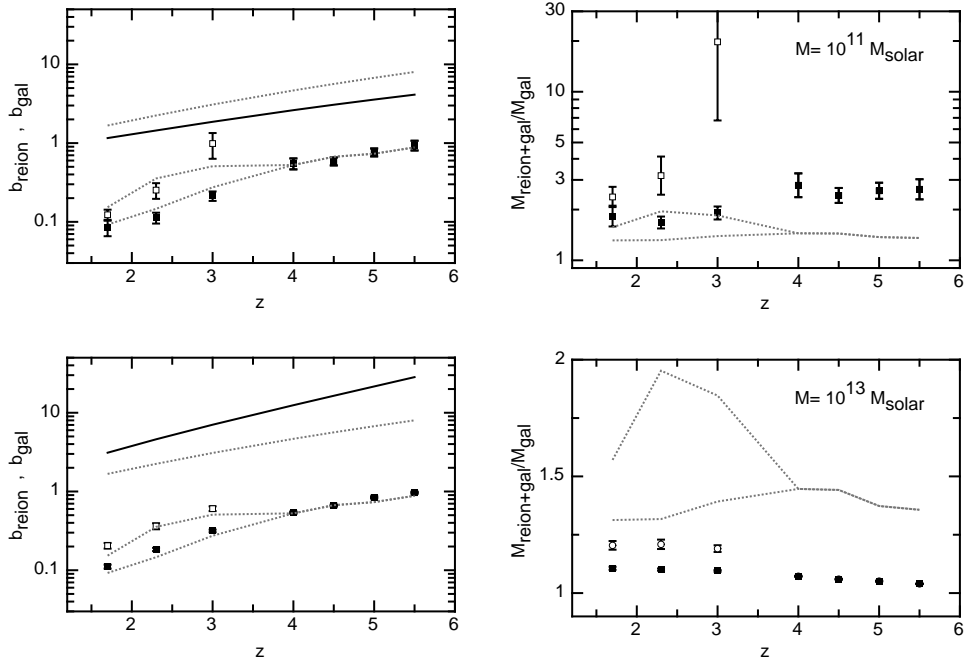}   
\caption{ Examples of clustering bias in Ly-break galaxies induced by reionization as a function of the halo mass. \textit{Left Hand Panels:} The bias introduced by reionization in cases where helium reionization at $z\sim3.5$ is considered in addition to hydrogen reionization (open squares) and where it is not (solid squares). The bias was computed assuming a flux evaluated at a rest-frame wavelength of $1350\AA$ within a $400\AA$ window. The galaxy bias is shown by the solid line for comparison. The error bars represent the statistical noise in the simulations due to the finite number of merger trees. \textit{Right Hand Panels:}  The factor by which the mass will be overestimated in clustering analyses where reionization is not considered. Results are shown for two halo masses, $M=10^{11}M_\odot$, and $M=10^{13}M_\odot$. In each case the corresponding results for $M=10^{12}M_\odot$, as presented in Figure~\ref{fig2}, are shown for comparison (light lines).}  
\label{fig3}  
\end{figure*}

We now present an estimate for the reionization induced bias in the  
sample of Ly-break galaxies. We show results at a scale of $R=10$  
comoving Mpc, which is a factor of $\sim3$ larger than the  
clustering length at $z\sim3$ (Adelberger et al.~2005). We evaluate  
the bias for fluxes measured at a rest frame wavelength of 1350\AA, and  
within a 400\AA~ wide band (this choice allows us to compare the predicted bias over the redshift range $1.7<z<5.5$).  In the upper right panel of  
Figure~\ref{fig2} we show the bias introduced by reionization in cases  
where hydrogen reionization alone is considered (solid squares), as  
well as cases where helium reionization at $z\sim3.5$ is considered in  
addition to hydrogen reionization (open squares). Also shown for  
comparison is the galaxy bias due to enhanced structure formation  
(solid line). In this figure the error bars represent the statistical  
noise in the simulations due to the finite number of merger trees. In  
order to better see the relative contributions of enhanced structure  
formation and reionization induced galaxy bias, in the central right  
panel we show the ratio of the component of bias introduced through  
reionization to the usual galaxy bias. Reionization represents a  
10-20\% correction to the galaxy bias in Ly-break galaxy samples at  
$1.7\la z\la3$. This correction corresponds to a predicted amplitude for  
the galaxy correlation function that can be 50\% larger than the prediction  
in the absence of consideration of reionization. Thus reionization  
provides a correction to the clustering amplitude that is  
in excess of the observational error for the existing  
Ly-break galaxy samples at $1.7\la z\la3$.  
  
One of the primary uses for measurements of clustering in a galaxy  
sample is the estimation of host halo mass. This mass estimate is made  
by measuring the bias, which is then interpreted theoretically in  
terms of host mass. However the results summarized in Figure~\ref{fig2} suggest that existing estimates of the galaxy bias could be  
systematically in error, at a level significantly larger than the  
observational error, due to the neglect of the effect of  
reionization. This in turn implies that estimates of the host masses  
in galaxy samples are also systematically in error. To evaluate the  
importance of this systematic error, we estimate the ratio of the  
inferred host masses with and without the inclusion of reionization,  
yielding  
\begin{equation}  
\ln(M_{\rm reion+gal})-\ln(M_{\rm gal}) \approx \frac{d\ln(M)}{db} [(b_{\rm  
g}+b_{\rm reion})-b_{\rm g}],  
\end{equation}  
or  
\begin{equation}  
\frac{M_{\rm reion+gal}}{M_{\rm gal}} \approx \exp{\left[b_{\rm  
reion}\frac{d\ln M}{db}\right]},  
\end{equation}  
where $dM/db$ is evaluated via equation~(\ref{bias}). The factor by  
which the mass will be overestimated in clustering analyses where  
reionization is not considered is plotted in the lower right hand  
panel of Figure~\ref{fig2}. We find that masses in existing Ly-break  
galaxy surveys (Adelberger et al.~2005) have been overestimated by  
factors of between 1.5 and 2.  
  
In addition to showing results evaluated at redshifts corresponding to  
the Ly-break galaxy sample, we also show results for hypothetical galaxy  
samples $4\leq z\leq5.5$. At these redshifts helium is not doubly  
reionized, and so all modifications to the star formation history are  
due to the reionization of hydrogen at $z\sim6$. This is in contrast  
to samples at $z\la3$ where helium reionization is  
complete. Figure~\ref{fig2} demonstrates that we expect to  
see a significant jump in the amplitude of the clustering for galaxy  
samples of fixed absolute magnitude following the double reionization  
of helium at $z\sim3.5$.

Up to this point we have presented results for $R=10$ comoving Mpc. We do  
not explicitly show results corresponding to other length scales in this  
paper for the following reason. On the scales of interest ($\sim3-100$  
cMpc), the variance is approximately a power-law with $R$, while  
the mass function is approximately a power-law with $\log(M)$. It turns out  
that these power-laws approximately cancel, leaving the bias induced by  
reionization almost independent of scale. This independence is a  
coincidence. A different slope of the primordial power-spectrum would have  
led to a scale dependent bias. However the conclusion that the bias is not  
scale dependent should be treated with caution for two reasons.  First, we  
are unable to rule out scale dependence at the level of a few percent at  
the numerical accuracy of our simulations. Second, as discussed in  
\S~\ref{scaldependance}, the scale-independence might be broken by  
additional astrophysical effects. Thus, future observations aiming to  
measure clustering at the percent level over a large range of spatial  
scales will need to carefully account for this possibility.  
  
\subsection{Mass dependence of the reionization bias}  
  
\begin{figure}  
\includegraphics[width=8.5cm]{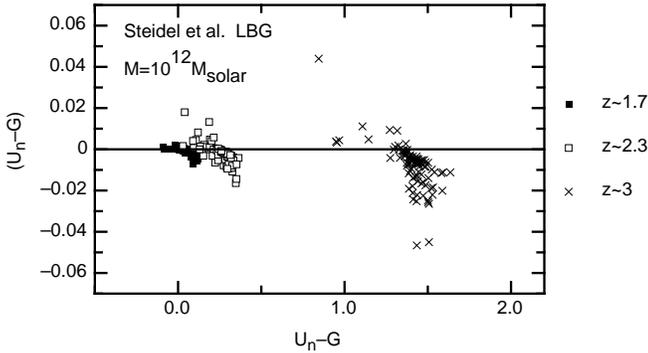}   
\caption{ Examples of how reionization will effect the observed colors of Ly-break galaxies. The figure shows the change in $U_n-G$ color for LBGs at $z\sim1.7$ (solid points), $z\sim2.3$ (open points) and $z\sim3$ (crosses), due to a fluctuation in the reionization redshift of $\delta z=0.25$.  }  
\label{fig4}  
\end{figure}   
  
Thus far our discussion of Ly-break galaxies has assumed a halo mass  
of $10^{12}M_\odot$, corresponding to observed Ly-break galaxies. In  
this section we describe the dependence of the predicted reionization  
induced galaxy bias on the halo mass. In Figure~\ref{fig3} we show  
examples of the clustering bias in Ly-break galaxies induced by  
reionization for halo masses of $M=10^{11}M_\odot$ and  
$10^{13}M_\odot$. In the left hand panels we show the bias introduced  
by reionization in cases where helium reionization at $z\sim3.5$ is  
considered in addition to hydrogen reionization (open squares) 
and where it is not (solid squares). As before the bias was  
computed assuming a flux evaluated at a rest-frame wavelength of $1350\AA$ within a $400\AA$  
window. The usual galaxy bias is shown by the solid line for  
comparison. In addition, in each case the corresponding results for  
$M=10^{12}M_\odot$, as presented in Figure~\ref{fig2}, are shown for  
comparison (light lines).   
  
We find that the contribution to the bias due to reionization is  
fairly insensitive to the halo mass. To understand this we note that  
although we would expect the larger halos to have begun forming  
earlier, and so to have their star formation histories less effected by the reionization of the IGM, this is offset by the steeper  
mass function of massive halos. On the other hand the galaxy bias due  
to enhanced structure formation in over dense regions is quite  
sensitive to the halo mass, and so we find that the fractional contribution to the galaxy bias is smaller for more  
massive systems. As a result, the systematic error introduced into the  
estimate of halo mass from clustering amplitude is less serious for  
more massive systems. In the right hand panels of Figure~\ref{fig3}  
we show the factors by which the mass will be overestimated in  
clustering analyses where reionization is not considered. While halos  
with masses near $M\sim10^{11}$ would be incorrectly inferred by a  
factor that could be larger than 3, the systematic error on very  
massive systems of $M\sim10^{13}M_\odot$ would be at only a level of  
10s of percent. This implies that the reionization bias will become  
more important as future surveys begin to discover populations of less  
massive galaxies at high redshift.

\subsection{Reionization and the observed colors of Ly-break galaxies}

The reionization induced bias should be sensitive to the selection  
band. In the case of Ly-break galaxies, we would therefore expect that  
the clustering amplitude would be sensitive to the band in which the  
flux selection was performed. Alternatively, overdense regions would  
therefore be expected to have a slightly bluer population of  
galaxies.   
  
In Figure~\ref{fig4} we show the change in $U_n-G$ color for LBGs at
$z\sim1.7$ (solid points), $z\sim2.3$ (open points) and $z\sim3$ (crosses),
due to a fluctuation in the reionization redshift of $\delta
z=0.25$. Galaxies in overdense regions, have systematically bluer colors
due to their younger stellar populations. The example shown for Ly-break
galaxies has fluctuations in $\Delta(U_n-G)$ color at the $\sim0.01-0.02$
level given a fluctuation in the redshift of overlap amounting to $\delta
z=0.25$ redshift units. On a scale of 10 comoving Mpc, the fluctuation in
the overlap redshift around a mean of $z=6$ is $\langle (\delta
z)^2\rangle^{1/2}\sim0.6$ (from equation~\ref{scatter}). Hence the expected
color variation between overdense and under dense regions would be
$\Delta(U_n-G)\sim0.03-0.05$ magnitudes. This expected correlation between
galaxy color and overdensity would be evidence for the reionization induced
galaxy bias, and could be used to calibrate its effect empirically.

This systematic variation in color is much smaller than the range of  
colors in the observed samples. However high redshift samples are  
selected to be redder than a certain limit.  In practice one would  
therefore have to be careful that the systematically bluer colors did  
not bias the sample {\em against} finding galaxies in overdense  
regions. At the redshifts of LBGs, the shift of the Ly-break with  
redshift primarily effects the $U_n-G$ color. On the other hand  
reddening effects both the $U_n-G$ and $G-\mathcal{R}$ colors.  As a  
result, LBGs are selected to lie above a line with positive gradient  
in the $(U_n-G)-(G-\mathcal{R})$ color-color space. We note that  
like reddening, the reionization induced color change will be in both  
bands, and will therefore transform the position of the galaxy in  
color-color space in a direction parallel to the selection criteria  
for LBGs. As a result, we do not expect the reionization induced color change to introduce a bias through the survey selection criteria.

\section{Surveys for Baryonic Oscillations}  
  
\begin{figure*}  
\includegraphics[width=13cm]{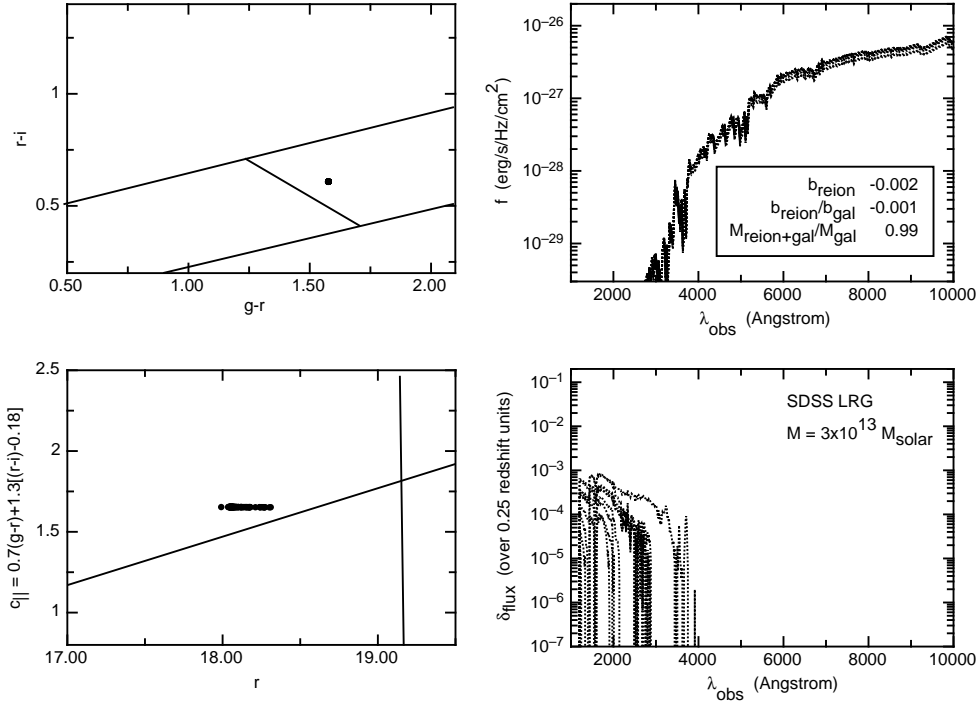}   
\caption{ The effect of reionization on the star formation histories of  
galaxies. These examples correspond to typical SDSS Luminous Red Galaxies  
(LRG) at $z=0.3$. \textit{Upper and Lower Left:} The primary color  
selection (Eisenstein et al.~2001) for LRGs, together with the locations of  
our model galaxies.  \textit{Upper Right:} Observed flux (including  
Ly$\alpha$ absorption) for ten example the Ly-break galaxies at $z=0.3$.  
\textit{Lower Right:} The magnitude change induced by a delay in  
reionization of 0.25 units of redshift. The corrections to the bias due to  
reionization are quoted in the upper right panel. We compute this bias  
using the rest-frame $r$-magnitude.}  
\label{fig5}  
\end{figure*}

\label{bao}  
  
We next apply our model to surveys that aim to measure baryonic acoustic oscillations in the clustering of  
galaxies at $z<1$. These surveys   
require exquisite accuracy of the clustering amplitude, and so the effect of reionization on galaxy bias could be particularly important. We consider two surveys, the existing SDSS Luminous Red Galaxy  
survey, and the planned WiggleZ survey.
  
\subsection{Luminous Red Galaxies}  
  
First we discuss the effect of reionization on the star formation histories  
of SDSS Luminous Red Galaxies (LRG) at $z=0.3$ (Eisenstein et al.~2001). By  
selection, LRGs are old galaxies with passively evolving stellar  
populations and no recent star formation. Thus, in order to model LRGs at  
$z\sim0.3$ we arbitrarily shut off star formation in the galaxies at  
$z=1$. The spectrum of the galaxy is not sensitive to the exact choice of  
redshift where star formation is curtailed, provided that the population of  
massive stars from the most recent star-burst have already died. However a  
cutoff in star formation is necessary if the models are to reproduce the  
correct colors of the observed sample.  
  
The upper and lower left hand panels of Figure~\ref{fig5} show the primary  
color selection\footnote{To estimate the colors of LRGs in this paper, we assume top-hat filters of central wavelength $\lambda_0$ and width $\Delta \lambda$ to approximate the Sloan Digital Sky Survey filter set. The filters have $(\lambda_0,\Delta\lambda)= (3543,564)$ for the $u$-band; $(\lambda_0,\Delta\lambda)=(4770,1388)$ for the $g$-band; $(\lambda_0,\Delta\lambda)=(6231,1372)$ for the $r$-band; $(\lambda_0,\Delta\lambda)=(7625,3524)$ for the $i$-band.}   
(Eisenstein et al.~2001) for LRGs at $z\sim0.3$. The model  
produces galaxies with the correct colors and observed flux, as is  
illustrated by the magnitudes and colors of the 100 modeled galaxies which  
are also plotted in the left hand panels of Figure~\ref{fig5}. In the upper  
right hand panel of Figure~\ref{fig5} we show examples of observed flux  
(including Ly$\alpha$ absorption) for ten model LRGs at $z=0.3$. These  
spectra show less variation than those of the Ly-break galaxies discussed  
in the previous section. This lack of variation is a feature of the LRG  
sample. Due to the lack of star formation in these galaxies, the spectra do  
not exhibit a sharp Ly-break. In the lower right panel of Figure~\ref{fig5}  
we show the fractional change in flux induced by a delay in reionization of  
0.25 units of redshift. We see that reionization has a very small effect on  
the observed flux of these galaxies. The resulting value of bias is quoted  
in the upper right panel. LRGs are selected to lie within a range of  
rest-frame absolute $r$-magnitudes, and we therefore calculate the bias at  
the rest-frame $r$-magnitude. Reionization will decrease the bias by $\sim  
0.1$\% in the LRG sample, and therefore the clustering amplitude by $\sim  
0.2$\%. Also shown is the fractional systematic error in the derived host  
mass ($\sim 1$\%).  
  
Eisenstein et al.~(2005) summarize the various corrections to the linear  
bias that have been previously considered when interpreting clustering  
data, including those due to non-linear gravity, and coupling of  
gravitational modes. On scales larger than $\sim40$ comoving Mpc, the sum  
of previously considered corrections drops below the 1\% level. For this  
reason among others, the correlation function of galaxies is considered to  
be a very clean tracer of the underlying large-scale mass-distribution, and  
in particular a perfect sample with which to investigate the baryonic  
oscillations in the matter power-spectrum (Eisenstein et al.~2005). It is  
therefore important to note that while the correction to the galaxy bias  
due to reionization predicted by our models is at a very low level for the  
LRG sample, it may nevertheless be comparable to the largest correction to  
linear theory yet described on the scales relevant to baryonic oscillation  
experiments. On the other hand, our model predicts no dependence of the  
reionization induced bias on scale. As a result it is very unlikely that  
the details of the reionization will adversely effect attempts to use the  
measurements of baryonic acoustic oscillations as a cosmic standard ruler  
(Blake \& Glazebrook~2003). We return to this point in \S~\ref{scaldependance}.

\subsection{Blue Star Forming Galaxies}  
  
\begin{figure*}  
\includegraphics[width=13cm]{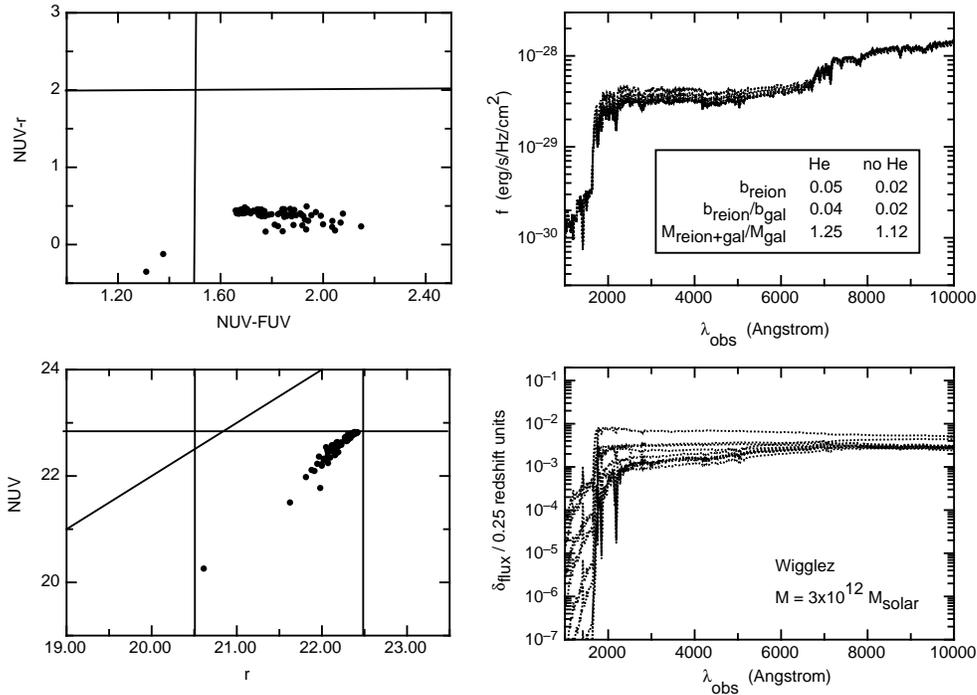}   
\caption{ The effect of reionization on the star formation histories of galaxies that will be selected by the WiggleZ Survey (Glazebrook et al.~2007) at $z=0.8$. \textit{Upper and Lower Left:} The primary color selection (Glazebrook et al.~2007) for Wiggles star forming galaxies at $z\sim0.8$, together with the points for our model galaxies.   
\textit{Upper Right:}  Observed flux (including Ly$\alpha$ absorption) for ten example the Ly-break galaxies at $z=0.8$.  \textit{Lower Right:} The magnitude change induced by a delay in reionization of 0.25 units of redshift. The corrections to the bias due to reionization are quoted in the upper right panel. This bias is computed at the observed $r$-band wavelength.}  
\label{fig6}  
\end{figure*}   
  
For our second example we consider the effect of reionization on the star
formation histories of galaxies that will be selected by the WiggleZ Survey
(Glazebrook et al.~2007) at $z=0.8$. Unlike the SDSS LRG sample considered
in the previous section, galaxies in the WiggleZ survey will be selected as
being star forming via the Ly-break using observations in the near and far
UV in addition to optical colors. In modeling these galaxies we therefore
do not impose a cutoff in the star formation prior to the observed
redshift.

The upper and lower left hand panels of Figure~\ref{fig6} show the primary
color selection\footnote{To estimate the UV colors in this paper, we assume
top-hat filters of central wavelength $\lambda_0$ and width $\Delta
\lambda$ to approximate the Galex filter set. The filters assumed have
$(\lambda_0,\Delta\lambda)= (1550,400)$ for the $FUV$-band; and
$(\lambda_0,\Delta\lambda)=(2500,1000)$ for the $NUV$-band.} (Glazebrook et
al.~2007) for WiggleZ star forming galaxies at $z\sim0.8$. As with the
previous examples the model produces galaxies with the correct optical and
UV colors as well as the correct observed fluxes. In the upper right panel
we show the observed spectra (including Ly$\alpha$ absorption) for ten
examples of WiggleZ selected galaxies at $z=0.8$.  In the lower right hand
panel of Figure~\ref{fig6} we show the fractional change in galaxy flux
induced by a delay in reionization of 0.25 units of redshift. Unlike the
LRG sample, the active star forming nature of the WiggleZ sample will mean
that (like the Ly-break galaxies at higher redshift) patchy reionization is
expected to significantly effect the observed clustering. The value of bias
due to reionization, the relative correction to the galaxy bias from
reionization, and the fractional change in the host mass inferred from the
clustering amplitude are quoted in the upper right panel of
Figure~\ref{fig6}. We quote results based both on models that include only
hydrogen reionization, and on models that also include the additional
heating of the IGM due to double reionization of helium.  The bias was
calculated at the observed $r$-band wavelength to be consistent with the
luminosity selection of the sample. Reionization will increase the bias by
$\sim 5$\% where helium reionization is included, and therefore the
clustering amplitude by $\sim 10$\%. The mass estimate would be inferred
incorrectly by a factor of as much as 25\% where reionization is ignored.

One of the underlying premises motivating galaxy surveys to measure the
baryonic acoustic oscillations is that the galaxies provide a nearly
perfect {\em geometrical} estimate of the distance, free from any
astrophysical complexities. However we have demonstrated that in the case
of star forming galaxies at moderate redshifts, the astrophysical effect of
reionization may enter the clustering statistics at the $\sim 5-10$\%
level. This level is significantly larger than the precision necessary for
measurement of the baryonic acoustic oscillations. Moreover, it is
important to note that the correction of $\sim5\%$ to the galaxy bias due
to reionization is the largest correction to linear theory yet described on
the scales relevant to baryonic oscillation experiments (Eisenstein et
al.~2005).  On the other hand, as mentioned earlier, our simple model
predicts that the bias due to reionization is, like the linear bias due to
enhanced formation in overdense regions, independent of scale. Thus in an
analysis that ignores reionization, the host mass would be misidentified,
but because the correction to the linear bias is not scale-dependent, the
unknown details of the reionization history may not compromise the
measurement the baryonic acoustic peak.

\section{Discussion}  
\label{discussion}

Before concluding we discuss several issues which arise from our
results and which will provide interesting areas for future research.

\subsection{Implications for the evolution of clustering in galaxy   
samples}

The observed spatial correlation function of galaxies can be used to
estimate the mass of the host dark-matter halo population through
comparison with theoretical calculations. Having determined this mass, the
evolution in the clustering of these galaxies can then also be computed and
compared with the clustering properties of different populations at later
times, with the aim of piecing together the evolution of the galaxy
population. Moreover having estimated the host halo mass, the predicted
number density of hosts can be compared with the observed number density of
objects in order to obtain the fraction of halos containing a galaxy of the
selected type at any one time.  By comparing the inferred mass of LBGs from
clustering data to the observed number counts, Adelberger et al.~(2005)
concluded that star formation in LBGs has a duty-cycle approaching unity.
This conclusion is consistent with our star formation model in which nearly
all model galaxies satisfy the LBG color selection criteria.
  
In this sub-section we consider the interpretation of LBG clustering  
evolution in light of the additional contribution to the observed galaxy  
bias from reionization.  The spatial correlation function of dark matter  
halos as a function of radius $r$ can be written in terms of the  
correlation function of dark-matter and the halo bias $b$ as  
\begin{equation}  
\xi_{\rm h}(r) = \xi_{\rm m}b^2(M),  
\end{equation}  
In practice this correlation function can be approximated using the   
parameterization  
\begin{equation}  
\label{approx}  
\xi_{\rm h} \approx \left(\frac{r}{r_0}\right)^{-\gamma},  
\end{equation}  
where $r_0$ is defined as the clustering length, and $\gamma\sim1.5$   
describes the observed clustering of galaxies. More biased samples have   
larger clustering lengths.

We have argued that reionization will increase the observed value of the
bias, by causing galaxies in overdense regions to have lower mass-to-light
ratios due to their younger stellar populations. Thus we also expect
reionization to increase the observed clustering length of a sample of
galaxies at fixed halo mass. As a result, neglect of reionization leads to
overestimation of the true clustering length for host dark-matter halos.
For small values of $b_{\rm reion}/b_{\rm gal}$, equation~(\ref{approx})
may be used to estimate the contribution to the observed clustering length
($\Delta r_{0,{\rm reion}}$) that results from reionization induced bias using the expression
\begin{eqnarray}  
\label{deltaR}  
\nonumber  
\Delta r_{0,{\rm reion}} &\approx & r_0\frac{2}{\gamma}\frac{b_{\rm   
reion}}{b_{\rm gal}}\approx 1.25 r_0\frac{b_{\rm reion}}{b_{\rm gal}}\\  
&\approx&1.4\left(\frac{r_0}{5.71}\right)\left(\frac{b_{\rm reion}/b_{\rm   
gal}}{0.2}\right),  
\end{eqnarray}  
where the units of length-scales in the latter equality are comoving Mpc.
  
The clustering evolution of LBGs was discussed by Adelberger et
al.~(2005). They measure the clustering length of LBGs at $z\sim1.7$,
$z\sim2.3$ and $z\sim3$, obtaining $r_0=5.7$, $r_0=6.0$ and $r_0=6.4$
comoving Mpc respectively, corresponding to halo masses of
$10^{12.1\pm0.2}M_\odot$, $10^{12\pm0.3}M_\odot$ and
$10^{11.5\pm0.3}M_\odot$. Using simulations, Adelberger et al.~(2005)
calculated the clustering lengths that these galaxies should have at lower
redshifts of $z\sim1$ and $z\sim0.2$, and then compared these evolved
clustering lengths to clustering studies of various populations of galaxies
from other surveys. In particular, Adelberger et al.~(2005) compared the
evolved clustering length for LBGs to galaxies in the DEEP Survey (Coil et
al.~2004), and in the Sloan Digital Sky Survey (Budavari et al.~2003).
Adelberger et al.~(2005) find that the LBG clustering length should evolve
to a value that is consistent with redder elliptical galaxies ($r_0\approx
9.4$ comoving Mpc at both $z=1$ and $z=0.2$), but which is larger than the
clustering length for both the whole DEEP galaxy sample at $z\sim1$
($r_0\approx 4.6$ comoving Mpc) and the blue Sloan Digital Sky Survey
(SDSS) galaxies at $z=0.2$ ($r_0=6.4$ comoving Mpc).
  
Based on these results Adelberger et al.~(2005) argued that the descendents
of LBGs will have clustering strengths that are significantly in excess of
typical galaxies in optical magnitude-limited surveys at low redshift, and
therefore that LBGs must have stopped forming stars before
$z\sim1$. However the results of this paper show that the clustering length
at $z\sim3$ has been overestimated by $\Delta r_{0,{\rm reion}}\sim1.5$
comoving Mpc.  Since the reionization induced bias decreases in influence
towards low redshift, and is small below $z=1$ (see following sections) we
conclude that, after accounting for the reionization induced bias, the
clustering of the hosts of LBGs may well be comparable to the blue
population of galaxies at $z<1$. Indeed, as shown in Figure~13 of
Adelberger et al.~(2005), the value of $\Delta r_{0,{\rm reion}}$ computed
for LBGs at $z\sim3$ is comparable to the difference in the clustering
length of normal ellipticals and normal blue galaxies in the Sloan Digital
Sky Survey at $z=0.2$.  Thus the effect of reionization on the observed
clustering of galaxies should be accounted for in studies that aim to link
galaxies at a range of epochs through the evolution of their clustering
properties.

\subsection{Helium reionization and scale dependent bias }  
\label{scaldependance}

The previous section (\S~\ref{bao}) ended with the positive suggestion that
reionization will not impact measurement of the baryonic acoustic peak in
samples of moderate redshift star-forming galaxies, due to the independence
of the reionization induced bias on scale. Before concluding this paper, we
describe an additional astrophysical situation which may compromise this
favorable conclusion.
  
Our simple model predicts the bias introduced through reionization to  
be independent of scale. However this model ignores several  
astrophysical effects that could introduce additional fluctuations in  
temperature within the IGM, and hence also introduce additional  
dependencies of the star formation history large scale  
overdensity. These additional fluctuations might include a scale  
dependence that is different to that of cosmic variance, and could  
therefore introduce a scale dependent component of reionization  
induced galaxy bias.  
  
For example, consider the epoch after HeIII overlap. At that time the  
mean-free-path for HeIII ionizing photons is limited by abundance and  
cross-section of Ly-limit systems, since the diffuse HeII has  
previously been ionized. During this epoch, heating of the IGM will be  
sourced by recombinations of HeIII ions. Now the recombination time is  
an order of magnitude longer than the Hubble time at the mean IGM  
density. However in the overdense regions containing filaments and  
sheets the recombination time would be shorter, and could approach a  
Hubble time in high density regions. As a result, while regions of low  
overdensity would cool adiabatically by the cosmic expansion, heating  
due to photo-ionization of HeII could be substantial in the overdense  
regions. This effect would introduce temperature fluctuations inside  
overdense regions of the IGM on scales larger than the  
mean-free-path.  
  
Thus, it is possible that the ionizing photon mean-free-path introduces a
length scale below which reionization induced bias is independent of scale,
but above which the reionization induced bias is scale dependent. At
$z\sim3$ the ionizing photon mean-free-path is $\sim 100$ comoving Mpc
(e.g. Bolton \& Haehnelt~2006). This scale is uncomfortably close to the
scale of the baryonic acoustic peak, implying that careful account will
need to be taken of reionization induced bias in galaxy surveys that select
star forming galaxies.  A proper analysis of this possibility would require
full numerical modeling and is beyond the scope of the present paper.

\subsection{Improvements to the model}

In the future, our simple model could be improved in several ways.  Our
model predicts that galaxies in regions that are reionized earlier form a
larger fraction of their stellar mass at later times, implying that these
galaxies form a greater fraction of their stellar mass in more massive
halos. We have assumed that the star formation efficiency is independent of
halo mass. However if high redshift galaxies are subject to mass-dependent
feedback effects (such as supernova feedback), then the star formation
history would be altered. The presence of feedback in low mass halos would
result in a larger fraction of the final stellar mass being formed after
reionization, and hence in an increase in the sensitivity of the final
mass-to-light ratio to the local reionization redshift. In addition, one
could incorporate metal enrichment of the star formation. Stars forming in
galaxies within overdense regions where reionization occurs early have
their star formation, and hence their metal enrichment, delayed. As we have
discussed, the resulting stellar populations are therefore observed to be
younger at a low redshift $z$. Since there is a delay in the enrichment of
the IGM following a burst of star formation, the younger stellar
populations will have slightly lower metallicity. Thus the metallicity of
populations observed at $z$ should be slightly dependent on the
reionization history. We expect that the lower metallicity in the younger
populations would tend to reduce the magnitude of the reionization induced
bias, since more highly enriched populations are bluer, and have lower
mass-to-light ratios (though this would be a second order effect). 
On the other hand, since we have not included metal
enrichment, our model also underestimates the variation between the UV
fluxes of stellar populations with different ages, and hence also
underestimates the contribution of reionization to the galaxy bias. In
addition to metallicities, one might also attempt to include the effects of
dust, which leads to larger extinction in younger galaxies (Shapley et
al. 2001).  This would redden the spectra of galaxies in regions that were
reionized at earlier times.

\section{Conclusions}  
  
\label{conclusions}

We have developed a model to estimate the effect of reionization on the
clustering properties of galaxy samples at intermediate redshifts.  Current
models of the reionization of the intergalactic medium predict that
overdense regions will be reionized early due to the presence of galaxy
bias. The IGM in these regions is heated through the absorption of the
ionizing radiation. The heating leads to an increased Jeans mass, and so
reionization suppresses the formation of low-mass galaxies. The suppression
of low mass galaxy formation in turn delays the build up of stellar mass in
the progenitors of massive low redshift galaxies. As a result of this
delayed buildup, the stellar populations observed in galaxies at later
times are on average slightly younger in overdense large-scale regions of
the Universe. Stellar populations fade as they age and so the resulting age
difference would lead to a lower mass-to-light ratio for galaxies in
overdense regions. In volume limited surveys, such as those now being
employed for large scale clustering studies, a fixed observed flux
threshold therefore contains lower mass galaxies (on average) in overdense
regions with a corresponding increase in the galaxy number density.
  
We have parameterized the reionization induced increase of the observed galaxy 
density in overdense regions in analogy with the traditional galaxy  
bias. Our modeling uses merger trees combined with a stellar synthesis  
code. We have used this model to demonstrate that reionization can have a  
significant and detectable effect on the clustering properties of galaxy  
samples that are selected based on their star-formation activity.  
  
In existing samples of Ly-break galaxies, the bias correction for
reionization is at the level of 10-20\%, leading to correction factors
between 1.5--2 in the mass inferred from clustering amplitudes. This effect
is present in existing samples of Ly-break galaxies at $1\la z\la3$
(Steidel et al.~2003), and provides a systematic correction to existing
analyses that is in excess of the statistical errors (Adelberger et
al.~2005). For example the reionization induced bias qualitatively changes
the conclusion of Adelberger et al.~(2005) that Ly-break galaxies stop
forming stars at $z\ga1$, and evolve into red elliptical galaxies by
$z\sim2$. Rather, allowing for reionization induced bias implies that
Ly-break galaxies could evolve into the blue populations observed at low
redshift with clustering lengths that are smaller than the massive red
galaxy population.
  
The reionization of helium, and the associated additional heating of the
IGM may lead to a sharp increase in the amplitude of the correlation
function of $\sim50\%$ for galaxies at fixed luminosity in the redshift
range $3\la z\la 4$.  Our model predicts that the reionization introduced
bias is approximately independent of scale. However we are unable to rule
out scale dependence at the level of a few percent due to the limited
numerical accuracy of our calculations. Further astrophysical complexities
not addressed in our model could alter this conclusion. Future experiments
aimed at measuring galaxy clustering at a precision level of a few percent
over a large range of spatial scales (with a goal of constraining the
initial conditions from inflation or the nature of dark matter and dark
energy), will need to carefully account for this possibility.
  
We find that the contribution to the bias due to reionization is fairly
insensitive to halo mass. This is in contrast to the galaxy bias from
enhanced structure formation in overdense regions which is a function of
halo mass. Hence the fractional contribution to the galaxy bias by
reionization is smaller for more massive systems, and as a result the
systematic error introduced into the estimate of halo mass from clustering
amplitude is less serious for more massive systems. We find that while the
reionization bias is already effecting clustering studies of Ly-break
galaxies, it will become even more important as future surveys begin to
discover populations of less massive high redshift galaxies.

The reionization induced bias should be sensitive to the type of selected
galaxies.  In the case of Ly-break galaxies, we would therefore expect that
the clustering amplitude would depend on the band in which the flux
selection was performed. Alternatively, overdense regions would therefore
be expected to have a slightly bluer population of galaxies. For Ly-break
galaxies our model predicts the systematic offset in $U_n-G$ color to be
$\sim0.03-0.05$ magnitudes. Note that this offset refers to the correlation
between color and large scale overdensity within a color-selected sample,
and not to the galaxy population on average. The average galaxy population
could show a different behavior. For example, galaxies in overdense regions
are observed at low redshift to be redder because they formed earlier, or
because their cold gas was heated by mergers or stripped by the hot IGM in
clusters.  A correlation between galaxy color and overdensity within a
Ly-break galaxy sample would be evidence for the reionization induced
galaxy bias, and could be used to calibrate its effect empirically.
  
Finally, we considered the importance of reionization induced bias for
current and upcoming surveys attempting to detect baryonic acoustic
oscillations. We find that the contribution to the bias from reionization
is very small in surveys of old stellar population galaxies at $z<1$, with
corrections of $\la 1\%$.  We also find that reionization should not impact
on measurement of the baryonic acoustic peak in samples of moderate
redshift star-forming galaxies, due to the independence of the reionization
induced bias on scale. However it is possible that the mean-free-path of
ionizing photons introduces a length scale below which reionization induced
bias is scale dependent, but above of which the reionization induced bias
is scale dependent. The scale of the mean-free-path is uncomfortably close
to the scale of the baryonic acoustic peak, implying that careful account
will need to be taken of reionization induced bias in galaxy surveys that
select star forming galaxies.

{\bf Acknowledgments} We thank Dan Stark for helpful comments on an early  
draft of this paper. This work was supported by the Australian Research  
Council (JSBW) and Harvard University grants (AL). JSBW acknowledges the  
hospitality of the Institute of Astronomy at Cambridge University where  
part of this work was undertaken.  
  
\newcommand{\noopsort}[1]{}

\begin{appendix}

\section{The dependence of the growth factor on large scale overdensity}  
\label{app1}  
  
In \S~\ref{model} we used the fact that the details of the merger tree
history are not sensitive to the local large-scale overdensity. In this
Appendix we demonstrate this independence.
  
Consider the conditional probability function for the number of progenitors
of mass between $M^\prime$ and $M^\prime+dM^\prime$ that a halo of mass $M$
breaks into when one takes a small redshift step $dz$
\begin{eqnarray}  
\label{dNdM}  
\nonumber \frac{dN}{dM^\prime} &=& -\sqrt{\frac{2}{\pi}}\frac{M}{M^\prime} \left(\sigma_{\rm cm}^2(M^\prime)-\sigma_{\rm cm}^2(M)\right)^{-1.5} \\  
&\times&\frac{d\delta_{\rm crit}}{dz} \left(\sigma_{\rm cm}^2(M^\prime)\frac{d\log \sigma_{\rm cm}}{dM}(M^\prime)\right) dz.  
\end{eqnarray}     
Equation~(\ref{dNdM}) is written in terms of comoving densities and
variances (labeled with sub-script "cm"). This equation could be modified
in the presence of a large scale overdensity, which we denote $\delta_{\rm
cm}$ when extrapolated to $z=0$ (making it ``comoving''), by adjusting the
overdensity for collapse as is done to derive galaxy bias in the
Press-Schechter formalism. This adjustment takes place in the term
\begin{eqnarray}  
\label{ddeltacdz}  
\nonumber \frac{d\delta_{\rm crit}}{dz} &=& \frac{\left(\delta_{\rm
c}\frac{D(\delta_{\rm cm},0)}{D(\delta_{\rm cm},z+dz)}-\delta_{\rm
cm}\right)-\left(\delta_{\rm c}\frac{D(\delta_{\rm cm},0)}{D(\delta_{\rm
cm},z)}-\delta_{\rm cm}\right)}{dz}\\ &=& \frac{\delta_{\rm
c}\frac{D(\delta_{\rm cm},0)}{D(\delta_{\rm cm},z+dz)}-\delta_{\rm
c}\frac{D(\delta_{\rm cm},0)}{D(\delta_{\rm cm},z)}}{dz},
\end{eqnarray}  
where we have explicitly included both the dependence of the growth factor  
$D$ on the comoving overdensity, and its normalization at $z=0$.  
  
There remains a possible dependence on large-scale overdensity in  
$d\delta_{\rm crit}/dz$ through the dependence on the growth factor.  
However the variation of the evolution in the growth factor due to  
$\delta_{\rm cm}$ is present both in the growth factor and in its  
normalization, so that the ratio $D(\delta_{\rm cm},0)/D(\delta_{\rm  
cm},z)$ should be independent of $\delta_{\rm cm}$ to leading order.  
  
To demonstrate this we compute the case of a zero cosmological constant (as
is appropriate at high redshift).  Locally, we can model a region of large
scale overdensity $\delta_{\rm cm}$ as being carved out of a universe with
a density parameter $\Omega_m(1+\delta_{\rm cm})$, where $\Omega_m$ is the
mean density parameter of our Universe. We can then compute the evolution
of structure and the local relation between scale factor and time, within
this modified universe. However when we observe galaxies in the overdense
region we see them at a redshift that is determined by the expansion
history of the actual Universe, not of the overdense region. To compare the
evolution of structure in the mean universe and overdense regions, we
therefore need to consider the growth factor within the overdense region at
a fixed time in the past (corresponding to the expansion of the actual
universe), rather than at a fixed scale factor. Furthermore, the growth
factor computed must then also be normalized by the growth factor in the
over dense region at the current time in the usual way. We have
\begin{equation}  
D \propto [\Omega_m (1+\delta_{\rm cm})]^{-1} a(\delta_{\rm cm}).  
\end{equation}  

Here $a(\delta_{\rm cm})$ is the overdensity dependent scale factor, which
is related to the age of the universe at $a(\delta_{\rm cm})$ through
\begin{equation}  
t = 2/3 H_0^{-1} [\Omega_m(1+\delta_{\rm cm})]^{-1/2} [a(\delta_{\rm
cm})]^{3/2},
\end{equation}  
where $H_0$ is the local value of Hubble's constant.  Substituting, we find
\begin{equation}  
D \propto H_0^{2/3} (1+\delta_{\rm cm})^{-2/3} t^{2/3},  
\end{equation}   
and hence   
\begin{equation}  
D(t)/D(t_0) = (t/t_0)^{2/3}.  
\end{equation}  
Thus the growth factor, normalized to the growth factor at the present day
is independent of $\delta_{\rm cm}$ (when the cosmological constant is
ignored).
  
The above arguments imply that since $\delta_{\rm cm}$ cancels in
equation~(\ref{ddeltacdz}), equation~(\ref{dNdM}), and therefore a merger
tree is insensitive to the large scale overdensity. While structure forms
earlier in overdense regions leading to an enhanced merger rate, there are
more halos with which to merge so that the merger rate per halo remains
constant.

\end{appendix}

\label{lastpage}  
\end{document}